\documentclass[prl,twocolumn,amsmath,amssymb,superscriptaddress,letterpaper]{revtex4}
\usepackage{graphicx}
\usepackage{bm}

\begin{document}

\title{Anomalous double peak structure in Nb/Ni superconductor/ferromagnet tunneling DOS}

\author{P. SanGiorgio}
\affiliation{Department of Physics, Stanford University,
Stanford, CA 94305}
\author{S. Reymond}
\affiliation{Ecole Polytechnique F\'{e}d\'{e}rale de Lausanne, Lausanne, Switzerland}
\author{M.R. Beasley}
\affiliation{Department of Applied Physics, Stanford University,
Stanford, CA 94305}
\author{J.H. Kwon}
\affiliation{Center for Strongly Correlated Materials Research,
Department of Physics and Astronomy, Seoul National University,
Seoul 151-742, KOREA}
\author{K. Char}
\email{kchar@phya.snu.ac.kr}
\affiliation{Center for Strongly Correlated Materials Research,
Department of Physics and Astronomy, Seoul National University,
Seoul 151-742, KOREA}

\date{\today}

\begin{abstract}
We have experimentally investigated the density of states (DOS) in Nb/Ni (S/F) bilayers as a
function of Ni thickness, $d_F$.  Our thinnest samples show the usual DOS peak at
$\pm\Delta_0$, whereas intermediate-thickness samples have an anomalous ``double-peak''
structure.  For thicker samples ($d_F \geq 3.5$~nm), we see an ``inverted'' DOS which has
previously only been reported in superconductor/weak-ferromagnet structures. We analyze the
data using the self-consistent non-linear Usadel equation and find that we are able to
quantitatively fit the features at $\pm\Delta_0$ if we include a large amount of spin-orbit
scattering in the model. Interestingly, we are unable to reproduce the sub-gap structure
through the addition of any parameter(s). Therefore, the observed anomalous sub-gap structure
represents new physics beyond that contained in the present Usadel theory.
\end{abstract}


\maketitle


The co-existence of superconductivity and ferromagnetism was first proposed by
Fulde and Ferrell~\cite{FULDE64} and Larkin and Ovchinnikov~\cite{LARKIN65} more than
forty years ago.  While some unusual materials have since been found with both superconducting
and ferromagnetic transitions (\emph{e.g.} ErRh$_4$B$_4$~\cite{MAPLE82}), much recent
interest has focused on conventional superconductor/ferromagnet (S/F) proximity effect
multi-layer systems.  A wide variety of unusual phenomena has been proposed for these
systems including oscillating critical temperatures, $T_c$~\cite{RADOVIC91}, $\pi$-state
Josesphson junctions~\cite{BUZDIN82}, and long-ranged odd-frequency triplet
superconductivity~\cite{BERGERET01}.

Qualitative evidence for the first two of these effects is convincing, but definitive
quantitative agreement with theory has been problematic.  The evidence for triplet
superconductivity is less certain, although a recent report by Keizer
\emph{et al}~\cite{KEIZER06} provides tantalizing evidence for such an effect.  One
reason for the difficulty in achieving quantitative agreement with theory is the
proliferation of physical effects that now have been incorporated into the theory,
leading to a concomitant proliferation of fitting parameters, which makes discriminating
fits to limited data sets nearly impossible.

In order to obtain more discriminating data sets and to further explore the SF proximity
effect in the case of strong ferromagnets, we have undertaken superconducting tunneling
densities of state (DOS) measurements on Nb/Ni thin-film bilayers.  By varying the Ni
thickness, $d_F$, we can track the spatial evolution of the behavior of the Cooper pairs
diffusing into the ferromagnet. This approach gives us much more information per sample
(the entire DOS spectrum) than $T_c$ or $J_c$ measurements, and is less sensitive to variations
in boundary parameters.  Analyzing these results with the most complete forms of the Usadel
theory available has allowed us to discriminate critically for the first time the relative
importance of the various physical effects now incorporated into the theory.  We find that
by far the most important parameter beyond the exchange field, $E_{ex}$, is the degree of
spin-orbit  scattering (first suggested by Demler \emph{et al}~\cite{DEMLER97}).  In addition,
we find an anomalous double-peak structure in the DOS that has not been reported previously
and that we have been unable to account for theoretically.

\begin{figure}
\includegraphics{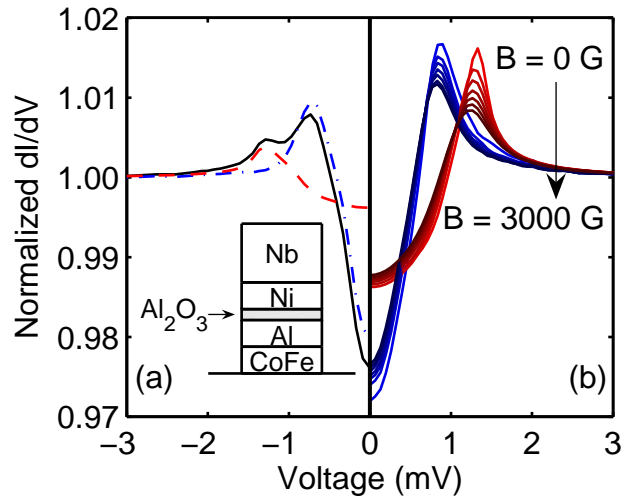}
\caption{\label{fig:isobfield} (a) Normalized conductance of the $d_F = 2.0$~nm sample, split into
\emph{outer-gap} and \emph{sub-gap} contributions.  The outer-gap contribution (dashed line),
a scaled version of the $d_F = 1.0$~nm curve and the sub-gap contribution (dash-dot line) sum to
the total conductance (solid line). (b) Normalized conductance of the $d_F = 1.5$~nm sample taken
in a perpendicular magnetic field, split into outer and sub-gap contributions.  The sub-gap peak shifts
in energy and changes amplitude slightly, but it does not broaden in an applied field, like the
outer-gap.}
\end{figure}

We use planar tunnel junctions of the form \emph{normal-insulator-ferromagnet-superconductor}.
A schematic of our sample geometry is shown in the inset of Fig.~\ref{fig:isobfield}. The
deposition of our samples and characterization of the tunnel  junctions has been documented
elsewhere~\cite{KIM05,REYMOND06}.  In brief, the various layers are sputtered and patterned
with stencil masks \emph{in situ} in a DC magnetron sputtering chamber without breaking
vacuum.  The Al$_2$O$_3$ tunnel barriers are formed by oxidizing the Al underlayer, and are
canonical in their behavior, except for the zero-bias anomalies commonly observed in tunnel
junctions incorporating magnetic materials; we discuss these below. We use a
Co$_{60}$Fe$_{40}$ backing of the Al electrode so as to reduce its critical temperature below
our lowest measurement temperature, which ensures normal/superconductor tunneling. We vary
$d_F$ from $0$~nm to $5$~nm in $0.5$~nm increments. Ni has a Curie Temperature of roughly
$600$~K, and should be ferromagnetic for film thicknesses greater than two atomic layers
($5$~\AA)~\cite{NEUGEBAUER59,BERGMANN78}. We have taken care to ensure that our Ni
films are ferromagnetic by measuring the superconducting resistive transition of the bilayer
in a parallel magnetic field: we detect hysteretic signals confirming the presence of
magnetism in all our films with $d_F \geq 1.0$~nm.


Tunneling measurements were performed at $0.28$~K using standard lock-in techniques.
Samples were measured in zero magnetic field and also in a perpendicular field
greater than $H_{c2}$.  In all of our samples, the normal-state conductance is a
V-shaped curve in which the zero-bias conductance (ZBC) is roughly $1\%-2\%$ lower
than the conductance at $5$~mV, which is typical of magnetic tunnel
junctions~\cite{LECLAIR01}. Since there are no systematic trends in the size of the
background, we do not consider it relevant to the superconducting DOS.  As
discussed in more detail in Ref.~\cite{REYMOND06}, we remove this background
conductance in our data analysis by dividing the zero-field conductance by the high-field
conductance, thus isolating the superconducting DOS. Using this procedure, we find
that the resultant DOS satisfies the sum rule on the total density of states, except
in our thickest samples, i.e. those with the smallest conductance variations.

\begin{figure}
\includegraphics{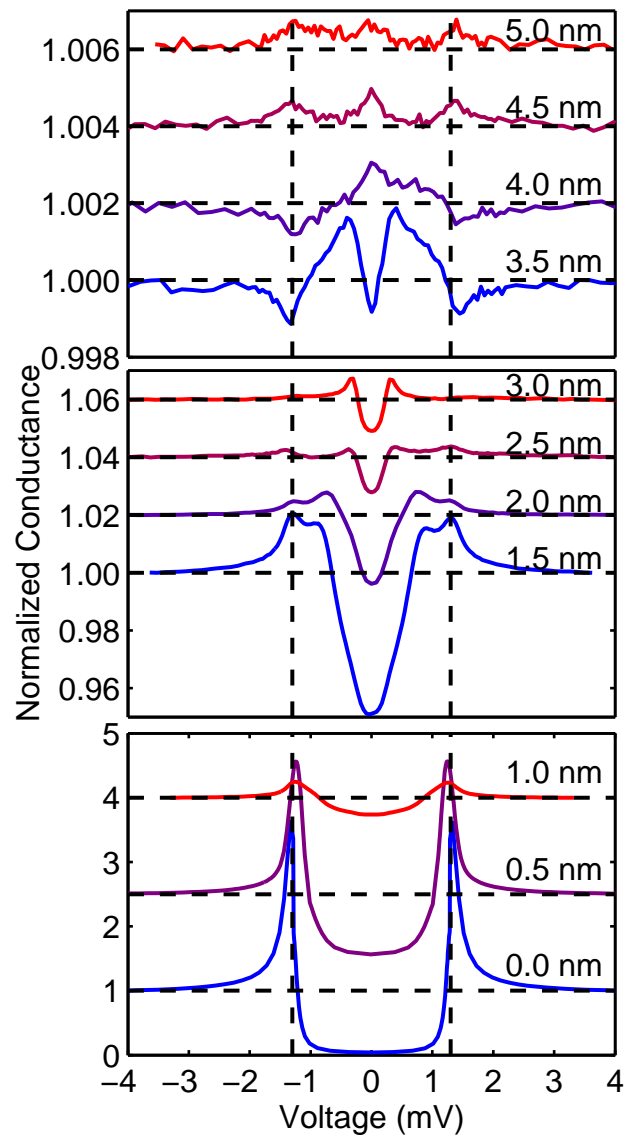}
\caption{\label{fig:didv} Normalized conductances taken at $0.28$~K for various Ni
thickness indicated inside. From the bottom to the top plot the vertical scale is
successively amplified. The curves are shifted for clarity. The vertical lines show that
the outside peaks energy remains unchanged from 0.5 to 4.5~nm.}
\end{figure}

The results of this normalization procedure for all measured $d_F$ are shown in
Fig.~\ref{fig:didv}, with $d_F = 0$ (tunneling into pure Nb) at the bottom and $d_F = 5.0$~nm
at the top, where the conductance scale has been magnified by roughly a factor of 1000 in comparison.
The $d_F = 0$ curve has two clear BCS coherence peaks at $\pm\Delta_0 = 1.3$~meV;
above that, we see that the addition of just $1.0$~nm of Ni increases the ZBC significantly.
In addition to the general trend of decreasing feature size as a function of $d_F$, for $d_F =1.5$~nm
we see a striking new feature: two coherence peaks on either side of zero-bias.
Continuing up Fig.~\ref{fig:didv}, we see this ``double-peak'' structure of the DOS for
$d_F = 1.5 - 3.0$~nm, with the interior peaks moving to lower voltages as $d_F$ increases.
At $d_F = 3.5$~nm, though, we see a qualitatively different DOS: the coherence peaks at
$V = \pm \Delta_0$ have ``inverted'' and are now conductance minima, while a narrow interior
gap remains. This inverted DOS was seen in the first reported measurement of the DOS in
S/F bilayers~\cite{KONTOS01}, yet has not been seen in other reports~\cite{CRETINON05,REYMOND06}.
The outer peak remains inverted at $d_F = 4.0$~nm, but by $d_F = 4.5, 5.0$~nm we have returned
to the non-inverted DOS with the addition of a very narrow peak at zero-bias.  These
measurements of thicker samples ($d_F \geq 4.5$~nm) must be taken with a grain of salt as
the feature size is quite small. It is our opinion that this zero-bias peak, which -- like all
other features of the DOS -- does not change in a small parallel or perpendicular magnetic
field is due to the steep voltage dependence of the background conductance and is therefore
a by-product of our normalization procedure.


We model our system in the dirty limit, which is applicable only when the elastic scattering
time, $\tau_e$, is shorter than all other relevant time-scales.  In the F-layer,
$\hbar / \tau_e \approx 400$~meV (from resistivity data), which is much greater than
either the estimated exchange field, $E_{ex} = 78$~meV (estimated from the Curie
Temperature~\cite{LIECHTENSTEIN86}), or the superconducting gap, $\Delta = 1.3~$meV.  In
this limit, the superconducting order parameter should obey the Usadel
equation~\cite{USADEL70,GUSAKOVA06},
\begin{multline}
-\frac{\hbar D}{2}\frac{\partial^2 \theta_{\uparrow(\downarrow)}}{\partial x^2} +
	\left(- i \omega \pm iE_{ex} + 2\Gamma_Z \cos\theta_{\uparrow(\downarrow)}\right)
	\sin\theta_{\uparrow(\downarrow)} + \\
	\Gamma_X \sin(\theta_{\uparrow}+\theta_{\downarrow}) \pm 
	\Gamma_{SO} \sin(\theta_{\uparrow}-\theta_{\downarrow}) =
	\Delta \cos\theta_{\uparrow(\downarrow)},
\label{eq:usadel}
\end{multline}
where $\theta_{\uparrow(\downarrow)}$ corresponds to the up (down) electron
density, $D$ is the diffusion constant, $\omega$ is the energy, $E_{ex}$ is the
exchange field, $\Gamma_{Z}$ and $\Gamma_{X}$ are the strength of the magnetic
scattering parallel and perpendicular to the quantization axis, and $\Gamma_{SO}$
is the strength of the spin-orbit scattering.  This equation is valid in the S-layer
when all magnetic terms ($E_{ex}, \Gamma_{Z}, \Gamma_{X}$) are zero and in the F layer
when the gap, $\Delta$, is zero.  $\Delta$ must also obey the self-consistent equation,
which at zero temperature is
\begin{equation}
\Delta(x) = \lambda\int_0^{\omega_D}d\omega(1/2)\text{Im}\left[\sin\theta_{\uparrow}
	+\sin\theta_{\downarrow}\right],
\label{eq:delta}
\end{equation}
where $\omega_D$ is the Debye frequency, and $\lambda$ is the coupling constant; this
can be solved iteratively.

The S-layer and F-layer occupy the space $-d_S < x < 0$ and $0 < x < d_F$, respectively.  We
supplement our equation with the usual (non-magnetic) boundary conditions: 
\begin{gather}
\frac{d\theta^{F}_{\uparrow(\downarrow)}}{dx}\Big|_{d_F}= 0,
-\xi_N \gamma_B \frac{d\theta^{F}_{\uparrow(\downarrow)}}{dx}\Big|_{0^+}
=\sin\left(\theta^{S}_{\uparrow(\downarrow)}-\theta^{F}_{\uparrow(\downarrow)}\right)
\label{eq:fbc}\\
\frac{d\theta^{S}_{\uparrow(\downarrow)}}{dx}\Big|_{-d_S}= 0,
-\xi_S \frac{\gamma_B}{\gamma}\frac{d\theta^{S}_{\uparrow(\downarrow)}}{dx}\Big|_{0^-}
=\sin\left(\theta^{S}_{\uparrow(\downarrow)}-\theta^{F}_{\uparrow(\downarrow)}\right)
\label{eq:sbc}
\end{gather}
 where $\xi_N = \sqrt{(\hbar D_F / 2 \Delta)}$, $\xi_S = \sqrt{(\hbar D_S / 2 \Delta)}$,
$\gamma_B = R_B A / \rho_F \xi_N$,  and $\gamma = \rho_S \xi_S / \rho_F \xi_N$.
This notation is modeled after Refs.~\cite{FOMINOV02,GUSAKOVA06}, but with
minor differences for ease of calculation.  Finally, the total DOS measured by our
junctions, $N(\omega)$, is
\begin{equation}
N(\omega) = (1/2)\text{Re}\left[\cos\theta_{\uparrow}^{F}(\omega)
	+ \cos\theta_{\downarrow}^{F}(\omega)\right]_{x = d_F}.
\label{eq:dosdef}
\end{equation}

If  we numerically solve the Usadel equation with $E_{ex} \gg \Delta$ and no scattering,
we find the characteristic decay and oscillation of the DOS as a function of $d_F$~\cite{BUZDIN00},
but as $d_F$ increases, it goes from \emph{inverted} to \emph{normal}, to \emph{inverted}, etc.
In other words, we expect the inverted --- not the normal --- DOS to be seen for thin
($0.2$~nm $\lesssim d_F  \lesssim 5$~nm) F layers.  Cottet \emph{et al}~\cite{COTTET05} were
the first to comment on this unusual prediction --- which has not been seen in any tunneling study
of S/F systems --- and suggested that spin-dependent interfacial phase shifts (SDIPS) could
resolve this disagreement.

We find that we can qualitatively account for the observed behavior through the addition of
either $\Gamma_{Z}$, $\Gamma_{SO}$, or SDIPS --- the quantitative fits discussed below favor
$\Gamma_{SO}$ as the dominant scattering term.  Equally important, as noted before, we find that
once the Usadel equation in the F-layer is essentially linear ($d_F \approx \xi_F$) the only
effect of increasing $d_F$ is to scale the DOS at the interface~\cite{KONTOS04,REYMOND06}.  This
means that none of these parameters will produce any sub-gap structure -- nor, for that matter,
will any other parameter in Eq.~\ref{eq:usadel}~\footnote{After discussion with A. Cottet, a
newer model of SDIPS was created~\cite{COTTET07}, which although promising does not produce
any sub-gap structure for large $d_S / \xi_S$.}

Thus, we are motivated to split empirically our DOS into two parts: a DOS with a single peak (or
inverted peak) at $\pm \Delta_0$, which we call the outer gap, and whatever remains, which we call
the sub gap. Since the $d_F = 1.0$~nm curve only has peaks at $\pm\Delta_0$ we can use it as a
template to isolate the outer-gap features of the other curves.  We want to break down the other
curves into a sum of two curves: one that is a scaled (in the conductance-axis) copy of the template
and another which will contain all of the sub-gap features.  In order to determine the size of the
different contributions, we adjust the scaling of the template such that the remaining sub-gap
contribution is as smooth as possible at $\pm\Delta_0$.  Figure~\ref{fig:isobfield}~(a) shows the
results of this process on the $d_F = 2.0$~nm sample.  Qualitatively, the anomalous contribution for
$1.5 \leq d_F \leq 3.5$~nm looks like a superconducting gap whose width decreases with increasing
$d_F$ and which stays un-inverted even when outer gap inverts.

\begin{figure}
\includegraphics{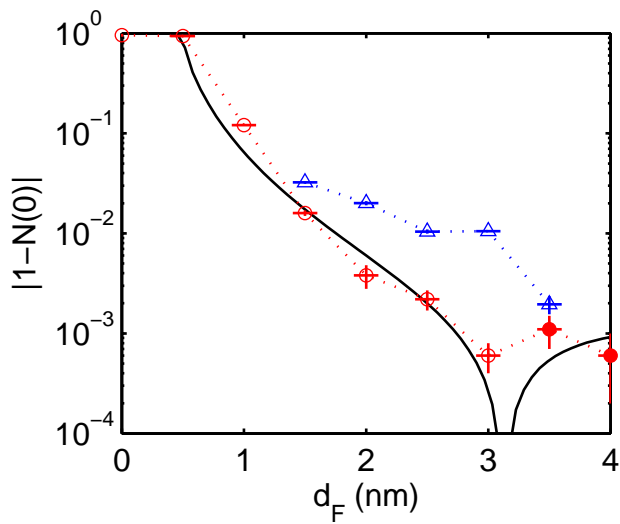}
\caption{\label{fig:zbc} Abs$(1 - N(0))$, the zero-bias conductance (ZBC), split into
\emph{outer-gap} (circles) and \emph{sub-gap} (triangles) contributions.  Open (filled) markers
represent a ZBC less (greater) than unity. The outer-gap contribution is fitted to Eq.~\ref{eq:usadel}
with $E_{ex} = 63$~meV and $\Gamma_{SO} = 46$~meV (thin line).}
\end{figure}

Once we have isolated the outer-gap contribution for all $d_F$, we can analyze it quantitatively
with the Usadel equation.  As noted above, a combination of $E_{ex}$, SDIPS, $\Gamma_{Z}$, and
$\Gamma_{SO}$ could be used, but we find that the best and simplest fits require only $E_{ex}$ and
$\Gamma_{SO}$. Figure~\ref{fig:zbc} shows the absolute  value of $1 - N(0)$, the ZBC, versus $d_F$.
The circles represent the ZBC of the outer-gap contribution and the triangles represent the sub-gap
contribution.  The line is the calculated ZBC from the Usadel equation with parameters
$E_{ex} = 63$~meV, $\Gamma_{SO} = 46$~meV, $\gamma_B = 0.54$, and $\gamma = 0.52$, which fits the
normal ZBC reasonably well~\footnote{Applying a similar analysis to our previously published
results on Nb/CoFe produces a reasonable fit for $E_{ex} = 109$~meV and $\Gamma_{SO} = 85$~meV. The
large value of $\Gamma_{SO}$ is necessary to account for the lack of an inversion.}.  Thus, we
conclude that the outer gap is well-described by the Usadel equation with an exchange field and
spin-orbit scattering.

We measured the $d_F = 1.5$~nm sample in a perpendicular magnetic field from $B = 0$ to $3000$~G
in $500$~G increments.  To isolate the outer gap, we measured the template curve ($d_F = 1.0$~nm)
at the same fields and used those as templates for each applied field.  Figure~\ref{fig:isobfield}~(b)
shows the results of this process: the outer curves are scaled versions of the template while the
inner curves are the remaining sub-gap feature.  The relative weight of the template needed to isolate
each sub-gap curve is roughly constant, but the resulting sub-gap curves decrease in size with
increasing field.  Further, the shape of the sub-gap curve remains virtually unchanged in field,
while the outer-gap peaks display significant broadening.


The existence of sub-gap structure alone indicates physics beyond the standard
Usadel treatment.  It is interesting to speculate what that might be.  One
intriguing possibility is that it is related to triplet superconductivity, perhaps
in combination with some spin-flip scattering at the SF interface (see
Bergeret \emph{et al}~\cite{BERGERET05} and references therein).  While it is
clearly premature to make this case confidently, we note that the relative
insensitivity of the shape of the sub-gap peak to magnetic field and the apparent
slower decay of its associated ZBC as a function of $d_F$ are qualitatively
consistent with such a possibility.  Additional theory will be required to fully
assess this possibility.


In summary, we have measured the DOS of Nb/Ni bilayers as a function of Ni
thickness.  In addition to tunneling features which are well-described by the
Usadel equation with a strong exchange field and spin-orbit coupling, we have
also discovered a robust sub-gap structure which cannot be explained by the
conventional theory. By isolating this sub-gap contribution, we have shown
that it behaves in a qualitatively different way from the Usadel contribution
in a perpendicular magnetic field, which leads us to inquire whether it is the
result of $m = 1$ triplet correlations in the bilayers.

\acknowledgments

We thank A. Cottet, T. Kontos, and H.-Y. Choi for stimulating discussions.
The authors acknowledge the support of U.S. DOE, NSF, and KOSEF through CSCMR.

\end{document}